\documentclass{ws-procs9x6}
\usepackage{epsfig}
\usepackage{amssymb}
\usepackage{multirow}
\newcommand\decay{%
\;\rule[0.5ex]{0.4pt}{1ex}\hspace*{-4pt}\rightarrow}
\newcommand{\lsim}{\mbox{$\stackrel{\scriptstyle <}
{\scriptstyle \sim}$}}
\newcommand{\gsim}{\mbox{$\stackrel{\scriptstyle >}
{\scriptstyle \sim}$}}

\begin{document}

\title{The Discovery of Neutrino Masses\footnote{\uppercase{I}nvited talk at
the \uppercase{X} \uppercase{I}nternational \uppercase{W}orkshop on
\uppercase{M}ultiparticle \uppercase{P}roduction (\uppercase{CF} 2002),
\uppercase{C}rete,
\uppercase{G}reece, \uppercase{J}une 2002} } 
\author{Norbert Schmitz}

\address{Max-Planck-Institut f\"ur Physik\\
F\"ohringer Ring 6,\\ 
D-80805 M\"unchen\\ 
E-mail: nschmitz@mppmu.mpg.de}


\maketitle
\abstracts{ The recent observation of neutrino oscillations with atmospheric
and solar neutrinos, implying that neutrinos are not massless, is a discovery
of paramount importance for particle physics and particle astrophysics. This
invited lecture discusses --- hopefully in a way understandable also for the
non-expert --- the physics background and the results mainly from the two most
relevant experiments, Super-Kamiokande and SNO. It also addresses the
implications for possible neutrino mass spectra. We restrict the discussion to
three neutrino flavours ($\nu_e, \nu_{\mu}, \nu_{\tau}$), not mentioning a
possible sterile neutrino.}
\section{Introduction}
Until recently one of the fundamental questions in particle physics has
been as to whether neutrinos have a mass ($m_\nu > 0$, massive neutrinos)
or are exactly massless (like the photon). This question is directly
related to the more general question whether there is new physics beyond
the Standard Model (SM): In the minimal SM, neutrinos have fixed helicity,
always $H(\nu) = -1$ and $H(\overline{\nu}) = + 1$. This implies $m_\nu =
0$, since only massless particles can be eigenstates of the helicity
operator. $m_\nu > 0$ would therefore transcend the simple SM. Furthermore,
if $m_\nu$ is in the order of 1 - 10 eV, the relic neutrinos from the Big
Bang ($n_\nu \approx 340/{\rm cm}^3$) would noticeably contribute to the
dark matter in the universe.

Direct kinematic measurements of neutrino masses, using suitable decays,
have so far yielded only rather loose upper limits, the present best values
being \cite{hag02}
\begin{equation}
\label{e1}
\begin{array}{lll}
m (\nu_e) < 3 \:{\rm eV} & & ({\rm from \: tritium} \: \beta \: {\rm decay})\\[1ex]
m (\nu_{\mu}) < 190 \:{\rm keV} & (90 $\%$ {\rm CL}) &  ({\rm from} \: \pi^+ \: {\rm
decay})\\[1ex]
m (\nu_{\tau}) < 18.2 \:{\rm MeV} & (95 $\%$ {\rm CL}) & ({\rm from} \: \tau
\: {\rm decays}).
\end{array}
\end{equation}
Another and much more sensitive access to neutrino masses is provided by
neutrino oscillations \cite{kay02}. They allow, however, to measure only
differences of masses squared, $\delta m^2_{ij} \equiv m^2_i - m^2_j$,
rather than masses directly. For completeness we summarize briefly the most
relevant formulae for neutrino oscillations in the simplest case, namely in
the vacuum and for only two flavours ($\nu_a, \nu_b$), e.g. ($\nu_e,
\nu_{\mu}$) (two-flavour formalism). The generalization to three (or more)
flavours is straight-forward in principle, but somewhat more involved in
practice, unless special cases are considered, e.g. $m_1 \approx m_2 \ll
m_3$ \cite{kay02}.
\begin{figure}
\label{fig1}
\begin{center}
\epsfig{file=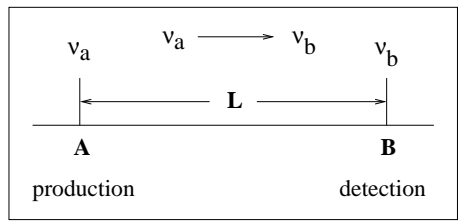,width=6cm}
\end{center}
\caption{Scheme of a neutrino oscillation experiment.}
\end{figure}

The two flavour eigenstates ($\nu_a, \nu_b$) are in general related to the two
mass eigenstates $(\nu_1, \nu_2)$ with masses ($m_1, m_2$) by a unitary
transformation: 
\begin{equation}
\label{e2}
 \left (
\begin{array}{c} \nu_a\\ \nu_b \end{array} \right ) = \left (
\begin{array}{c@{\hspace{0.3cm}}c} \cos \theta & \sin \theta\\ -\sin \theta &
\cos \theta \end{array} \right ) \cdot \left ( \begin{array}{c} \nu_1\\ \nu_2
\end{array} \right ) \end{equation} 
where $\theta$ is the mixing angle. If $m_1
\neq m_2$, the two mass eigenstates evolve differently in time, so that for
$\theta \neq 0$ the given original linear superposition of $\nu_1$ and $\nu_2$
changes with time into a different superposition. This means that flavour
transitions (oscillations) $\nu_a \to \nu_b$ and $\nu_b \to \nu_a$ can occur
with certain time-dependent oscillatory probabilities. In other words 
(Fig.~\ref{fig1}):  If a neutrino is produced (or detected) at A as a flavour
eigenstate $\nu_a$ (e.g.\ $\nu_{\mu}$ from $\pi^+ \to \mu^+ + \nu_{\mu}$), it is
detected, after travelling a distance (baseline) $L$, at B with a probability
$P (\nu_a \to \nu_b )$ as flavour eigenstate $\nu_b$ (e.g.\ $\nu_e$ in $\nu_e n
\to pe^-$). The transition probability $P (\nu_a \to \nu_b) = P
(\overline{\nu}_a \to \overline{\nu}_b) = P (\nu_b \to \nu_a)$ is given by
\begin{equation} \label{e3} \left.\begin{array}{ll} P (\nu_a \to \nu_b )& =
\sin^2 2\theta \cdot \sin^2 \left ( {\displaystyle\frac{\delta m^2}{4}} \cdot
{\displaystyle\frac{L}{E}}\right )\\[2.5ex] &= \sin^2 2\theta \cdot \sin^2
\left ( 1.267  {\displaystyle\frac{\delta m^2} {{\rm eV}^2} \cdot \frac{L/{\rm
m}}{E/{\rm MeV}}}\right ) \end{array}\right\rbrace \begin{array}{l} {\rm for}\
\nu_a \neq \nu_b\\ ({\rm flavour~change}\\ \nu_a \to \nu_b) \end{array}
\end{equation} 
\begin{displaymath} \hspace*{-3.3cm} P (\nu_a \to \nu_a ) = 1 -
P (\nu_a \to \nu_b )\ \ \ \ \ \mbox{(survival of $\nu_a$)} \end{displaymath}
where $\delta m^2 = m^2_2 - m^2_1$ and $E =$ neutrino energy. Thus the
probability oscillates when varying $L/E$, with $\theta$ determining the
amplitude $(\sin^2 2\theta )$ and $\delta m^2$ the frequency of the
oscillation. The smaller $\delta m^2$, the larger $L/E$ values are needed to
see oscillations, i.e.\ significant deviations of $P (\nu_a \to \nu_b )$  from
zero and of $P (\nu_a \to \nu_a )$ from unity. Notice the two necessary
conditions for $\nu$ oscillations: (a) $m_1 \neq m_2$ implying that not all
neutrinos are massless, and (b) non-conservation of the lepton-flavour
numbers. 

In (\ref{e3}), $L$ and $E$ are the variables of an experiment, and
$\theta$ and $\delta m^2$ the parameters (constants of Nature) to be
determined. The original situation ($P (\nu_a \to \nu_b ) = \delta_{ab}$) is
restored, if in (\ref{e3}) the distance $L$ is an integer multiple of the
oscillation length $L_{\rm osc}$ which is given by
\begin{equation}
\label{e4}
L_{\rm osc} = 4 \pi \dfrac{E}{\delta m^2} = 2.48 \dfrac{E/{\rm MeV}}{\delta
m^2/{\rm eV}^2} \, {\rm m}.
\end{equation}

The masses $m (\nu_a )$ and $m (\nu_b )$ of the flavour eigenstates are
 expectation values of the mass operator, i.e.\ linear combinations of $m_1$ and $m_2$:
\begin{equation}
\label{e5}
\begin{array}{l}
m (\nu_a ) = \cos^2 \theta \cdot m_1 + \sin^2 \theta \cdot m_2\\[1ex]
m (\nu_b )= \sin^2 \theta \cdot m_1 + \cos^2 \theta \cdot m_2.
\end{array}
\end{equation}

\section{Flavour change of atmospheric neutrinos}
The most convincing evidence for a flavour change of atmospheric neutrinos
was found in 1998 by Super-Kamiokande\cite{sk1,sk2}, after first
indications were observed by some earlier experiments (Kamiokande
\cite{fuk94}, IMB \cite{bec92}, Soudan 2 \cite{all99}).

Atmospheric neutrinos are created when a high-energy cosmic-ray proton (or
nucleus) from outer space collides with a nucleus in the earth's atmosphere,
leading to an extensive air shower (EAS) by  cascades of secondary
interactions. Such a shower contains many $\pi^\pm$ (and $K^\pm$) mesons (part
of) which decay according to 
\begin{equation} 
\label{e6}
\begin{array}{ll@{\hspace{1cm}}ll} \pi^+ , K^+ \rightarrow & \mu^+ \nu_{\mu} &
\pi^- , K^- \rightarrow & \mu^- \overline{\nu}_\mu \\ &\decay e^+ \nu_e
\overline{\nu}_\mu && \decay e^- \overline{\nu}_e \nu_{\mu} \, , \end{array}
\end{equation} 
yielding atmospheric neutrinos. From (\ref{e6}) one would expect
in an underground neutrino detector  a number ratio of 
\begin{equation}
\label{e7} 
\frac{\mu}{e} \equiv \frac{\nu_{\mu} + \overline{\nu}_\mu} {\nu_e +
\overline{\nu}_e} = 2\, , 
\end{equation} 
if all $\mu^\pm$ decayed before
reaching the detector. This is the case only at rather low shower energies
whereas with increasing energy more and more $\mu^\pm$ survive due to
relativistic time dilation and may reach the detector as background (atmospheric $\mu$). Consequently  the expected
$\mu /e$ ratio rises above 2 (fewer and fewer  $\nu_e , \overline{\nu}_e$) with
increasing $\nu$ energy. For quantitative predictions Monte Carlo (MC)
simulations, which include also other (small) $\nu$ sources, have been performed, using measured $\mu$ fluxes as
input, modelling the air showers in detail, and yielding the fluxes of the
various neutrino species ($\nu_e , \overline{\nu}_e , \nu_{\mu} ,
\overline{\nu}_\mu$) as a function of the $\nu$ energy \cite{hon95}. 

Atmospheric neutrinos reaching the underground Super-K
detector can be registered by neutrino reactions with nucleons inside the detector, the
simplest and most frequent reactions being CC quasi-elastic scatterings:
\begin{equation}
\label{e8}
({\rm a})\mbox{\hspace{0.3cm}}
\begin{array}{l}
\nu_e n \to p e^-\\
\overline{\nu}_e p \to ne^+
\end{array}
\mbox{\hspace{0.8cm} (b)\hspace{0.3cm}}
\begin{array}{l}
\nu_{\mu} n \to p \mu^-\\
\overline{\nu}_\mu p \to n\mu^+ \, .
\end{array}
\end{equation}
\begin{figure}
\begin{center}
\epsfig{file=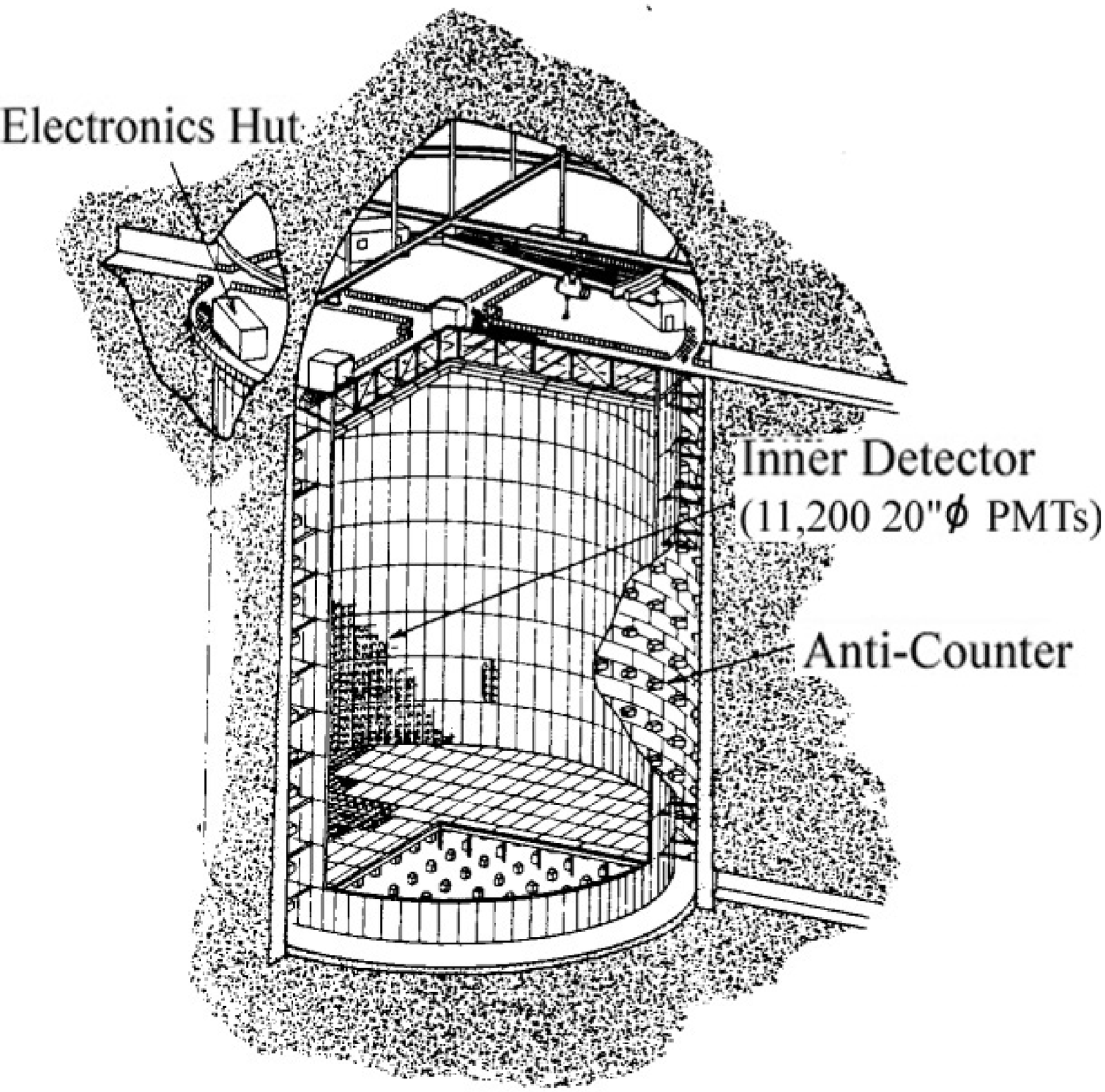,width=7.2cm}
\end{center}
\caption{Schematic view of Super-Kamiokande \protect\cite{nak94}.}
\label{fig2}
\end{figure}

Super-K
(Fig.~\ref{fig2})\cite{nak94}
is a big water-Cherenkov detector in the Ka\-mi\-o\-ka Mine (Japan) at a depth of
$\sim$ 1000 m. It consists of 50 ktons (50\,000 m$^3$) of 
ultrapurified water
in a cylindrical tank (diameter = 39 m, height = 41 m). The inner detector
volume of 32 ktons is watched by 11\,146 photomultiplier tubes (PMTs,
diameter = 20$''$) mounted on the volume's surface and providing a 40\%
surface coverage. The outer detector, which
tags entering particles and exiting particles,
 is a 2.5 m thick water layer
surrounding the inner volume and looked at by 1885 smaller PMTs (diameter =
8$''$). 
A high-velocity charged particle passing through the water produces a
cone of Cherenkov light which is registered by the PMTs. The Cherenkov
image of a particle starting and ending inside the inner detector is a
ring, the image of a particle starting inside
 and leaving the inner detector is
a disk. A distinction between an $e$-like event (\ref{e8}a) and a
$\mu$-like event (\ref{e8}b) is possible 
(with an efficiency of $\gsim\ 98\%$)
from the appearance of the image:
an $e^\pm$ has an image with a diffuse, fuzzy boundary 
whereas the boundary of a $\mu^\pm$ image is sharp. The
observed numbers of $\mu$-like and $e$-like events give directly the
observed $\nu$-flux
ratio $(\mu /e)_{\rm obs}$ (eq.~\ref{e7}) which is to be compared with the 
MC-predicted ratio $(\mu /e)_{\rm MC}$ (for no $\nu$ oscillations)
by computing the double ratio
\begin{equation}
\label{e9}
R = \frac{(\mu /e)_{\rm obs}}{(\mu /e)_{\rm MC}}\, .
\end{equation}
Agreement between observation and expectation implies $R = 1$.

The events are separated into fully contained events (FC, no track leaving the
inner volume, $\langle E_\nu \rangle \sim 1 \, {\rm GeV}$) and partially contained
events (PC, one or more tracks leaving the inner volume, $\langle E_\nu \rangle
\sim 10 \, {\rm GeV}$). For FC events the visible energy $E_{\rm vis}$, which is
obtained from the pulse heights in the PMTs, is close to the $\nu$ energy. With
this in mind, the FC sample is subdivided into sub-GeV events ($E_{\rm vis} <
1.33$ GeV) and multi-GeV events ($E_{\rm vis} > 1.33$ GeV). In the multi-GeV
range the $\nu$ direction can approximately be determined as the direction of
the Cherenkov-light cone, since at higher energies the directions of the
incoming $\nu$ and the outgoing charged lepton are close to each other.
\tabcolsep1.5mm
\begin{table}
\tbl{Results on the double-ratio {\it R}. The first error is
statistical, the second systematic (kty = kilotons $\cdot$ years).}
{\begin{tabular}{|l|l|ll|}
\hline
Super-K & $R = 0.652 \pm 0.019 \pm 0.051$ & sub-GeV & $(E_{\rm vis} < 1.33 \, {\rm
GeV})$\\ 
(70.5 kty) & $R = 0.661 \pm 0.034 \pm 0.079$ & multi-GeV &  $(E_{\rm
vis} > 1.33 \, {\rm GeV})$\\ 
\hline 
Soudan 2 & & &\\
(5.1 kty) & \raisebox{1.5ex}[-1.5ex]{$R = 0.68 \pm 0.11 \pm 0.06$} & &\\
\hline 
\end{tabular}}\label{t1}    
\end{table}

Recent results on {\it R} from Super-K \cite{sk2} and Soudan 2 \cite{man01} are
given in Tab.~\ref{t1}. All three {\it R} values are significantly smaller than
unity (``atmospheric neutrino anomaly'') which is due, as it
turns out (see below), to a deficit of $\nu_{\mu} , \overline{\nu}_\mu$ and
not to an excess of $\nu_e , \overline{\nu}_e$
in $(\mu /e)_{\rm obs}$. A natural explanation of this deficit is that
some $\nu_{\mu} , \overline{\nu}_\mu$ have oscillated into 
$(\nu_e , \overline{\nu}_e)$ or $(\nu_{\tau} , \overline{\nu}_\tau)$
according to (\ref{e3}) before reaching the detector. 

This explanation has become evident, with essentially only $\nu_{\mu} \rightarrow
\nu_{\tau}$ remaining (see below), by a study of the $\nu$ fluxes 
as a function of the zenith angle $\Theta$ between the vertical (zenith) and
the $\nu$ direction. A $\nu$ with  $\Theta \approx 0^\circ$ comes from above
(down-going $\nu$) after travelling a distance of $L\ \lsim\ 20$ km (effective
thickness of the atmosphere); a $\nu$ with $\Theta \approx 180^\circ$ reaches
the detector from below (up-going $\nu$) after traversing the whole earth with
$L \approx 13000$ km. 
\begin{figure}
\begin{center}
\epsfig{file=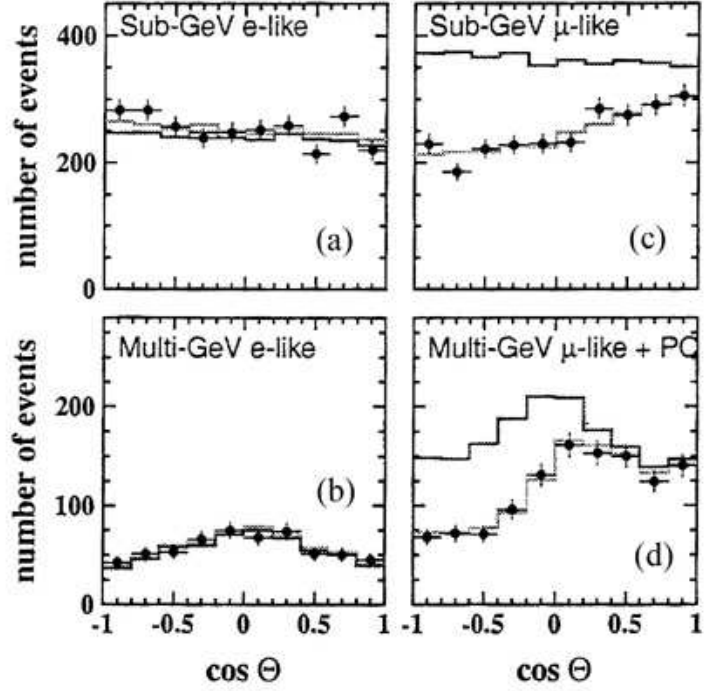,width=10cm}
\end{center}
\caption{Zenith-angle distribution of (a) sub-GeV {\it e}-like, (b) multi-GeV
{\it e}-like, (c) sub-GeV $\mu$-like, and (d) multi-GeV $\mu$-like + PC
events. (The PC events turned out to be practically all $\nu_{\mu}$ events). The
points show the data, the full histograms the MC predictions for no oscillations
and the dotted histograms the best fit by $\nu_{\mu} \rightarrow \nu_{\tau}$
oscillations. From Super-K\protect\cite{sk2}. }
\label{fig3}
\end{figure}

The zenith angular distributions (zenith angle of the charged lepton) as
measured by Super-K\cite{sk2} are shown in Fig. \ref{fig3} for {\it e}-like and
$\mu$-like events, in each event class separately for sub-GeV and multi-GeV
events. The full histograms show the MC predictions for no oscillations. The {\it
e}-like distributions (a) and (b) are both seen to be in good agreement with the
predictions which implies that there is no $\nu_e$ excess and no noticeable
$\nu_{\mu} \rightarrow \nu_e$ transition. The $\mu$-like distributions (c) and
(d) on the other hand both show a $\nu_{\mu}$ deficit with respect to the
predictions. For multi-GeV $\mu$-like events (d), for which the $\nu$ and $\mu$
directions are well correlated (see above), the deficit increases with
increasing zenith angle, i.e. increasing flight distance {\it L} of the $\nu$
between production and detection; it is absent for down-going muons ($\Theta
\approx 0^{\circ}$) and large for up-going muons ($\Theta > 90^{\circ}$). For
sub-GeV $\mu$-like events (c) the dependence of the deficit on $\Theta$ is much
weaker, owing to the only weak correlation between the $\nu$ and $\mu$
directions.

In conclusion, all four distributions of Fig. 3 are compatible with the
assumption, that part of the original $\nu_{\mu}$ change into $\nu_{\tau}$
(thus not affecting the {\it e}-like distributions), if their flight distance
{\it L} is sufficiently long ($L \, \gsim \, L_{\rm osc}$).

This conclusion is supported by a Super-K measurement of the zenith angular
distribution of up-going muons with $\Theta > 90^{\circ}$ that
enter the detector from outside~\cite{sk3,sk2}. Because of their large zenith angle they
cannot be atmospheric muons --- those would not range so far into the earth
---, but are rather produced in CC reactions by energetic up-going $\nu_{\mu},
\overline{\nu}_\mu$ in the rock surrounding the detector. A clear deficit is
observed for upward muons stopping in the detector ($\langle E_\nu \rangle \sim
10$ GeV) whereas it is much weaker for upward through-going muons ($\langle
E_\nu \rangle \sim 100$ GeV). A deficit of atmospheric $\nu_{\mu},
\overline{\nu}_\mu$ has also been observed by the MACRO
collaboration\cite{amb98} in the Gran Sasso Underground Laboratory in a similar
measurement, their ratio of the numbers of observed to expected events being
$\mu_{\rm obs}/ \mu_{\rm exp} = 0.72 \pm 0.13$ (three errors added in
quadrature) for upward through-going muons ($\langle E_\nu \rangle \sim 100$
GeV).

A two-flavour oscillation analysis, with $\sin^2 2 \theta$ and $\delta m^2$ as
free parameters, has been carried out by the Super-K collaboration, using their
data on (partially) contained events (Fig. 3) and including also their data on
up-going muons. A good fit with $\chi^2 /NDF = 135/152$ has been
obtained\cite{sk2} for $\nu_{\mu} \leftrightarrow \nu_{\tau}$, the best-fit
parameters being:
\begin{equation}
\label{e10}
\delta m^2 = 3.2 \cdot 10^{-3} \, {\rm eV}^2 \, , \, \sin^2 2 \theta = 1.
\end{equation}
\begin{figure}
\begin{center}
\epsfig{file=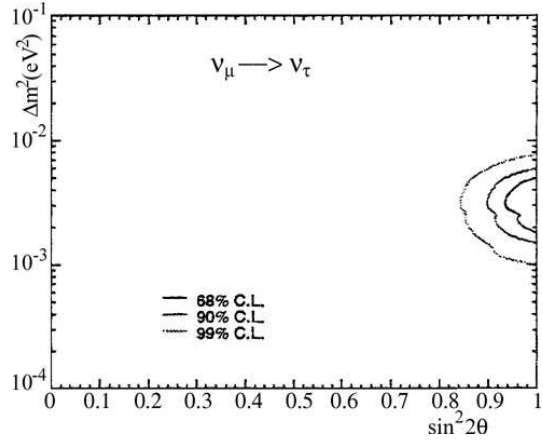,width=7cm}
\end{center}
\caption{Regions (to the right of the curves) allowed at 68 \%, 90 \% and 99 \% CL in the ($\sin^2 2 \theta,
\delta m^2$) plane for $\nu_{\mu} \leftrightarrow \nu_{\tau}$ oscillations. From
Super-K\protect\cite{sk2}.}
\label{fig4}
\end{figure}
Fig.~\ref{fig4} shows the allowed regions with 68 \%, 90 \% and 99 \% CL in the
parameter plane. The best fit is also shown by the dotted histograms in Fig.
\ref{fig3}, where excellent agreement with the data points is observed. From
(\ref{e4}) and (\ref{e10}) one obtains an oscillation length of $L_{\rm osc}=
775 \, {\rm km} \cdot E/{\rm GeV}$. Thus, a flavour-change signal is not
expected, because of $L \ll L_{\rm osc}$, (a) for neutrinos with $\Theta
\approx 0^{\circ}$ (i.e. $L \, \lsim \, 20$ km) and $E \, \gsim \, 1$ GeV (Fig.
\ref{fig3}d), and (b) for neutrinos producing upward through-going muons with
$E \sim 100$ GeV (see above) so that $L_{\rm osc} \sim 80000$ km --- much
larger than the diameter of the earth. 

No good fit could be obtained for $\nu_{\mu} \leftrightarrow \nu_e$ oscillations.
In addition, $\overline{\nu}_e$ disappearance ($\overline{\nu}_e \rightarrow
\overline{\nu}_{\rm X}$) has not been observed by two long-baseline reactor
experiments (CHOOZ\cite{apo99} and Palo Verde\cite{boe01}) with $L \approx 1$
km and $\langle E \rangle \sim 3$ MeV, which rule out $\delta m^2 > 0.7 \cdot
10^{-3} \, {\rm eV}^2$ for $\sin^2 2 \theta = 1$, and $\sin^2 2 \theta > 0.1$ for large $\delta m^2$.

In summary: Atmospheric neutrinos have yielded convincing evidence, mostly
contributed by Super-K, that $\nu_{\mu} \leftrightarrow \nu_{\tau}$ oscillations
take place with parameters given by Fig. 4 and Eq. (\ref{e10}). There is no other
hypothesis around that can explain the data. One therefore has to conclude that
not all neutrinos are massless.

\section{Flavour change of solar neutrinos} Very exciting discoveries regarding
neutrino masses have recently been made with solar neutrinos, in particular by
the Sudbury Neutrino Observatory (SNO). Solar neutrinos\cite{alt01} come from the fusion
reaction
\begin{equation}
\label{e11}
4p \to {\rm He}^4 + 2e^+ + 2\nu_e
\end{equation}
inside the sun with a total energy release of 26.7 MeV after two $e^+ e^-$
annihilations. The $\nu$ energy spectrum extends up to about 15 MeV with an
average of  $\langle E_\nu \rangle = 0.59$ MeV. The total $\nu$ flux from the
sun is $\phi_\nu = 1.87 \cdot 10^{38}$ s$^{-1}$ resulting in a flux density
of $6.6 \cdot 10^{10} \, {\rm cm}^{-2} \, {\rm s}^{-1}$ on earth.

Reaction (\ref{e11}) proceeds in various steps in the pp chain or CNO
cycle, the three most relevant out of eight different $\nu_e$ sources being: 
\begin{equation}
\label{e12}
\begin{array}{l@{\ :\ \ }l@{\hspace{0.3cm}}l}
pp& p + p \to D + e^+ + \nu_e &(E_\nu < 0.42\ {\rm MeV}, \, 0.91)\\[1ex]
{\rm Be}^7& {\rm Be}^7 + e^- \to {\rm Li}^7 + \nu_e & 
 (E_\nu = 0.86\ {\rm MeV}, \, 0.07)\\[1ex]
{\rm B}^8 & {\rm B}^8 \to
{\rm Be}^8 + e^+ + \nu_e & (E_\nu < 14.6\ {\rm MeV}, \, \sim 10^{-4}) .
\end{array}
\end{equation}
The second number in each bracket gives the fraction of the total solar $\nu$
flux. Energy spectra of the $\nu_e$ fluxes from the various sources and rates
for the various detection reactions have been predicted in the framework of the
Standard Solar Model \mbox{(SSM)\cite{bah89,bah01}.} With respect to these
predictions a $\nu_e$ deficit from the sun has been observed in the past by
various experiments as listed in Tab.~\ref{t2} (see ratios Result/SSM). These
deficits, the well-known ``solar neutrino problem'', could be explained by $\nu$
oscillations $\nu_e \to \nu_X$ into another flavour $X$ ($\nu_e$ disappearance)
either inside the sun (matter oscillations, Mikheyev-Smirnov-Wolfenstein (MSW)
effect\cite{mik86}) or on their way from sun to earth (vacuum oscillations, $L
\approx 1.5 \cdot 10^8$ km), see below.
\tabcolsep1.5mm
\begin{table}
\tbl{The five previous solar $\nu$ experiments and their results (adopting a recent
  compilation in Table 8 of Ref.\protect\cite{bah01}). The SSM is
  BP2000\protect\cite{bah01}.}
{\vspace*{0.3cm}
\begin{tabular}{|l|l|l|l|}
\hline
&&Threshold&Result\\
\raisebox{1.5ex}[-1.5ex]{Experiment}&
\raisebox{1.5ex}[-1.5ex]{Reaction}&
[MeV]&
(Result/SSM)\\
\hline\hline
&&&\\[-1ex]
Homestake\protect\cite{cle98} & Cl$^{37}$$(\nu_e , e^-)$Ar$^{37}$&
$E_\nu > 0.814$& 2.56 $\pm$ 0.23 SNU\\
&&&(0.34 $\pm 0.06$)\\[1ex]
GALLEX + GNO\protect\cite{bel01} & Ga$^{71}$$(\nu_e , e^-)$Ge$^{71}$&
$E_\nu > 0.233$& $74 \pm 7$  SNU$^+$\\
&&&(0.58 $\pm$ 0.07)\\[1ex]
SAGE\protect\cite{gav01} & Ga$^{71}$$(\nu_e , e^-)$Ge$^{71}$&
$E_\nu > 0.233$&75 $\pm$ 8 SNU$^+$\\
&&&(0.59 $\pm$ 0.07)\\[1ex]
Kamiokande\protect\cite{fuk96} & $\nu e \to \nu e$&
$E_\nu > 7.5$&(2.80 $\pm$ 0.38) $\cdot 10^6$ cm$^{-2}$ s$^{-1}$\\
&&&(0.55 $\pm$ 0.13)\\[1ex]
Super-Kamiokande\protect\cite{suz01} & $\nu e \to \nu e$&
$E_\nu > 5.5$&$\left (2.40{+0.09 \atop -0.08}\right ) 
\cdot 10^6$ cm$^{-2}$ s$^{-1}$\\
&&&(0.48 $\pm$ 0.09)\\
\hline
\end{tabular}}
\par\vspace{0.1cm}
{1 SNU (Solar Neutrino Unit)
= 1 $\nu_e$ capture per 10$^{36}$ target nuclei per sec}\\
$^+$ The latest values (Neutrino 2002) are $71 \pm 6$ for GALLEX + GNO and
$71^{+7}_{-8}$ for SAGE.
\label{t2}
\end{table}

\begin{figure}
\begin{center}
\epsfig{file=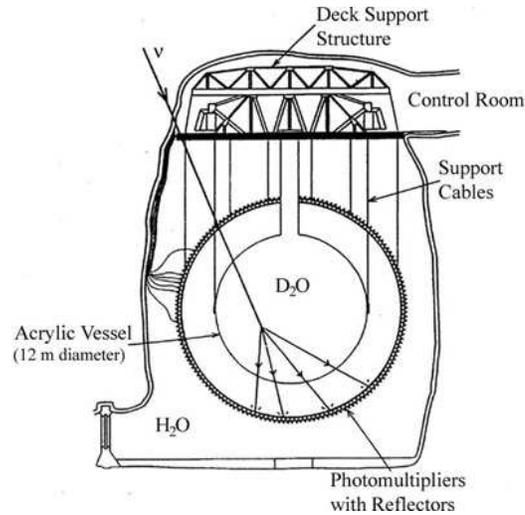,width=6.8cm}
\end{center}
\caption{Schematic drawing of the SNO detector.}
\label{fig5}
\end{figure}

We now discuss the new results from SNO\cite{ahm02,ahm12}. The SNO
detector\cite{bog00} (Fig. \ref{fig5}) is a water-Cherenkov detector, sited
2040 m underground in an active nickel mine near Sudbury (Canada). It comprises
1000 tons of ultra-pure heavy water (D$_2$O) in a spherical transparent acrylic
vessel (12 m diameter) serving as a target and Cherenkov radiator. Cherenkov
photons produced by electrons in the sphere are detected by 9456  20
cm-photomultiplier tubes (PMTs) which are mounted on a stainless steel
structure (17.8 m diameter) around the acrylic vessel. The vessel is immersed
in ultra-pure light water (H$_2$O) providing a shield against radioactivity
from the surrounding materials (PMTs, rock).

SNO detects the following three reactions induced by solar B$^8$-neutrinos above
an electron threshold of 5 MeV for the SNO analysis (d = deuteron):
\begin{equation}  
\label{e13}  
\begin{array}{lclll} 
\nu_e + {\rm d} &
\rightarrow &  \mbox{ \boldmath $e^-$} + p + p & {\rm (CC)} &
\hspace{3mm}E_{\rm thresh} = 1.44 \, {\rm MeV}\\ 
\nu_X + {\rm d} & \rightarrow &
\nu_X + p + n & {\rm (NC)} & \hspace{3mm}E_{\rm thresh} = 2.23 \, {\rm MeV}\\  
& &
\multicolumn{3}{l} {n + {\rm d} \rightarrow {{\rm H}^3} + \gamma \, (6.25 \, {\rm
MeV}), \, \gamma + e^- \rightarrow \gamma + \mbox{\boldmath $e^-$}} \\ 
\nu_X +
e^- & \rightarrow & \nu_X + \mbox{\boldmath $e^-$}  & {\rm (ES)} 
\end{array} 
\end{equation}
where the Cherenkov-detected electron is indicated by bold printing. The
charged-current ({\it cc}) reaction (CC) can be induced only by $\nu_e$ whereas the
neutral-current ({\it nc}) reaction (NC) is sensitive, with equal cross sections, to all
three neutrino flavours $\nu_e, \, \nu_{\mu}, \, \nu_{\tau}$. Also elastic
$\nu_e$-scattering (ES) is sensitive to all flavours, but with a cross section
relation 
\begin{equation}
\label{e14}
\sigma (\nu_{\mu} e) = \sigma (\nu_{\tau} e) = \varepsilon \sigma (\nu_e e)
\end{equation}
where $\varepsilon$ = 0.154 above 5 MeV according to the electroweak theory.
($\varepsilon \ne 1$ since $\nu_{\mu,\tau} e$ scattering goes only via {\it nc},
whereas $\nu_e e$ scattering has in addition to {\it nc} also a contribution from {\it cc}).

Data taking by SNO began in summer 1999. For each event (electron) the
effective kinetic energy {\it T}, the angle $\Theta_{\rm sun}$ with respect to
the direction from the sun, and the distance (radius) {\it R} from the detector
center were measured. The principle of the analysis goes
as follows: The three measured distributions $N(x)_{\rm meas}$ of $x = T, \,
\Theta_{\rm sun}, \, R^3$ from 2928 events with $5 < T < 20$ MeV can be fitted
by three linear combinations \begin{equation} \label{e15}  N(x) = N_{\rm CC}
\cdot w_{\rm CC} (x) + N_{\rm NC} \cdot w_{\rm NC} (x)  + N_{\rm ES} \cdot
w_{\rm ES} (x) +  N_{\rm BG} \cdot w_{\rm BG} (x) \end{equation} where $w_{\rm
i}(x)$ are characteristic probability density functions known from Monte Carlo
simulations (e.g. $w_{\rm ES} (\cos \Theta_{\rm sun}$) is strongly peaked in
the direction from the sun, i.e. towards $\cos \Theta_{\rm sun} = 1$), and the
parameters $N_{\rm i}$ are the numbers of events in the three categories
(\ref{e13}) to be determined from the fit. $N_{\rm BG}$
for the background was fixed as determined from a background calibration. A good
extended maximum likelihood fit to the measured distributions was obtained
yielding (errors symmetrized):
\begin{equation}
\label{e16}
N_{\rm CC} = 1967.7 \pm 61.4, \, N_{\rm NC} = 576.5 \pm 49.2, \, N_{\rm ES} =
263.6 \pm 26.0.
\end{equation}
From each of these event numbers $N_{\rm i}$ a B$^8$-neutrino flux
$\Phi^{\rm SNO}_{\rm i}$ was determined, using the known cross sections for
reactions (\ref{e13}) and the SSM B$^8$-$\nu$ spectrum. The exciting result
(in units of $10^6$ cm$^{-2}$ s$^{-1}$) is\cite{ahm02} (statistical and systematic errors added in quadrature):
\begin{equation}
\label{e17}
\Phi^{\rm SNO}_{\rm CC} = 1.76 \pm 0.10, \, \Phi^{\rm SNO}_{\rm NC} = 5.09 \pm
0.62, \, \Phi^{\rm SNO}_{\rm ES} = 2.39 \pm 0.26
\end{equation}
where $\Phi^{\rm SNO}_{\rm ES}$ has been computed using $\sigma(\nu_e e)$,
\mbox {i.e. assuming} no $\nu_e$-oscillations. $\Phi^{\rm SNO}_{\rm ES}$ agrees
nicely with the Super-K result\cite{fuk01}\ $\Phi^{\rm SK}_{\rm ES} = 2.32 \pm
0.09$, computed with the same assumption.

$ \Phi^{\rm SNO}_{\rm CC}$ is the genuine $\nu_e$ flux $\Phi (\nu_e)$ arriving
at earth. For the case that the $\nu_e$ created in the sun arrived at earth all
as $\nu_e$, i.e. there were no $\nu_e$ oscillations, one would expect
$\Phi_{\rm CC} = \Phi_{\rm NC} = \Phi_{\rm ES}$. The SNO result (\ref{e17})
shows that this is obviously not the case, i.e. that there is significant
direct evidence for a non-$\nu_e$ component in the solar $\nu$ flux arriving at
earth.

The two fluxes $\Phi(\nu_e)$ and $\Phi(\nu_{\mu \tau})$ (= $\nu_{\mu} +
\nu_{\tau}$ flux) and the total $\nu$ flux $\Phi_{\rm tot}$ have been determined
from (\ref{e17}) by a fit using the three relations
\begin{equation}
\label{e18}
\begin{array}{lcl}
\Phi_{\rm CC} & = & \Phi(\nu_e)\\
\Phi_{\rm NC} & = & \Phi(\nu_e) + \Phi(\nu_{\mu \tau}) = \Phi_{\rm tot}\\
\Phi_{\rm ES} & = & \Phi(\nu_e) + \varepsilon \Phi (\nu_{\mu \tau}) \,  {\rm
with} \;\varepsilon = 0.154
\end{array}
\end{equation}
with the result
\begin{equation}
\label{e19}
\Phi (\nu_e) = 1.76 \pm 0.10 \,\,\, {\rm and} \,\,\, \Phi(\nu_{\mu \tau}) = 3.41 \pm
0.65.
\end{equation}
Notice that $\Phi(\nu_{\mu \tau})$ is different from zero by 5.3 $\sigma$ which
is clear evidence for some ($\sim$ 66 \%) of the original $\nu_e$ having changed
their flavour. Furthermore, the measured value (\ref{e17}) $\Phi^{\rm SNO}_{\rm NC}
= \Phi^{\rm SNO}_{\rm tot} = 5.09 \pm 0.62$ (or the value $\Phi_{\rm tot} =
\Phi(\nu_e) + \Phi(\nu_{\mu \tau}) = 5.17 \pm 0.66$ from the fit result(\ref{e19}))
agrees nicely (within the large errors) with the SSM value\cite{bah01} $\Phi^{\rm
SSM}_{\rm tot} = 5.05^{+1.01}_{-0.81}$; this agreement is a triumph of the Standard Solar
Model.

\begin{figure}
\begin{center}
\epsfig{file=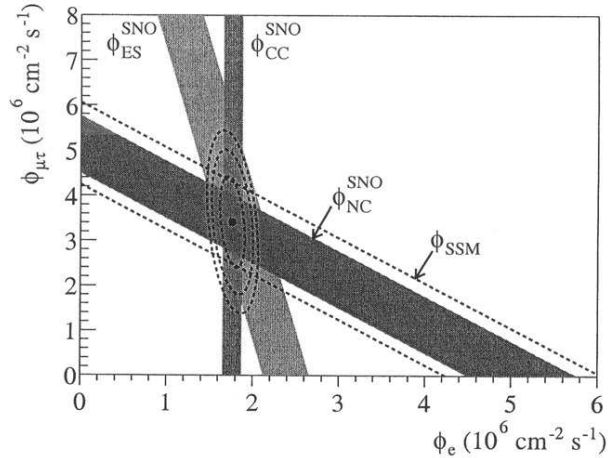,width=8cm}
\end{center}
\caption{Fluxes of B$^8$-neutrinos as determined by SNO\protect\cite{ahm02}. The bands
show the flux $\Phi_{\mu \tau}$ of ($\nu_{\mu} + \nu_{\tau})$ vs. the flux
$\Phi_e$ of $\nu_e$ according to each of the three experimental relations and
the SSM relation\protect\cite{bah01} in (\ref{e20}). The intercepts of these
bands with the axes represent the $\pm 1 \sigma$ errors. The point in the
intersection of the bands indicates the best-fit values (\ref{e19}). The
ellipses around this point represent the 68 \%, 95 \% and 99 \% joint
probability contours for $\Phi_e, \Phi_{\mu \tau}$. From Ref. \protect\cite{ahm02}.}
\label{fig6}
\end{figure}

The SNO analysis is summarized in the $\left [ \Phi(\nu_e), \, \Phi(\nu_{\mu
\tau}) \right ]$ plane, Fig.~\ref{fig6}. The four bands show the straight-line
relations (with their errors):
\begin{equation}
\label{e20}
\begin{array}{lcl}
\Phi^{\rm SNO}_{\rm CC} & = & \Phi(\nu_e) = 1.76 \pm 0.10\\[1ex]
\Phi^{\rm SNO}_{\rm ES} & = & \Phi(\nu_e) + 0.154 \cdot \Phi(\nu_{\mu \tau}) = 2.39 \pm
0.26\\[1ex]
\Phi^{\rm SNO}_{\rm NC} & = & \Phi(\nu_e) + \Phi(\nu_{\mu \tau}) = 5.09 \pm
0.62\\[1ex]
\Phi^{\rm SSM}_{\rm tot} & = & \Phi(\nu_e) + \Phi(\nu_{\mu \tau}) =
5.05^{+1.01}_{-0.81}.
\end{array}
\end{equation}
Full consistency of the three measurements (\ref{e17}) amongst themselves and with
the SSM is observed, the four bands having a common intersection.
\begin{table}
\tbl{Best-fit values for the five solutions from Ref.~29}
{\begin{tabular}{|lllll|}\hline 
\rule[2ex]{0mm}{1mm}
Solution && $\delta m^2$(eV$^2$) & $\tan^2 \theta$ & $\chi^2_{\rm min}$\\ [1ex]
\hline 
LMA \rule[2ex]{0mm}{1mm}&
\multirow{3}*{\hskip-8mm{\mbox{$\Big\rbrace$} MSW}} & $4.2 \times 10^{-5}$ & $2.6
\times 10^{-1}$  &29.0\\ 
SMA && $5.2 \times 10^{-6}$ & $5.5 \times 10^{-4}$ & 31.1\\
LOW   && $7.6 \times 10^{-8}$ & $7.2 \times 10^{-1}$ & 36.0\\
Just So$^2$ && $5.5 \times 10^{-12}$ & $1.0 \times 10^{0}$ & 36.1\\
VAC && $1.4 \times 10^{-10}$ & $3.8 \times 10^{-1}$ & 37.5\\
\hline
\end{tabular}}\label{t3}
\end{table}

\begin{figure}
\begin{center}
\epsfig{file=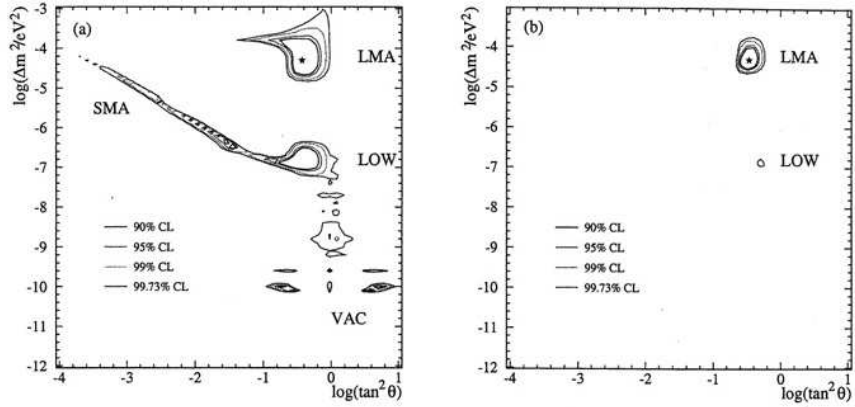,width=11.5cm}
\end{center}
\caption{Regions allowed at the indicated confidence levels in the parameter
plane as determined from a $\chi^2$ fit (a) to the SNO day and night energy spectra
alone, and (b) with the addition of data from the other solar experiments and
of SSM predictions~\protect\cite{bah01}. The star in the LMA solution indicates the best
fit (\ref{e21}). From SNO~\protect\cite{ahm12}.}
\label{fig7}
\end{figure}

A two-flavour oscillation analysis $(\nu_e \leftrightarrow \nu_{\mu} \,\, {\rm
or} \,\, \nu_{\tau})$ has been carried out by the SNO collaboration\cite{ahm12}.
Prior to SNO several global oscillation analyses were performed using all
available solar neutrino data, including the Super-K measurements of the
electron energy spectrum and of the day-night asymmetry (which could originate
from a regeneration of $\nu_e$ in the earth at night)\cite{fuk01,fuk11}. Five allowed
regions (e.g. at 3$\sigma$, i.e. 99.7 \% CL) in the ($\tan^2 \theta, \, \delta
m^2$) plane were identified, their best-fit values e.g. from Ref.\cite{bah11}
being listed in Table~\ref{t3}. These solutions, apart from Just So$^2$, were
also found by SNO\cite{ahm12} when only using their own data (measured day and
night energy spectra), Fig.~\ref{fig7}a. When including also the data from the
previous experiments as well as SSM predictions in their analysis, only the
large-mixing-angle (LMA) MSW solution is strongly favoured (Fig.~\ref{fig7}b),
the best-fit values being
\begin{equation}
\label{e21}
\delta m^2 = 5.0 \cdot 10^{-5} \, {\rm eV}^2, \, \tan^2 \theta = 0.34 \; (\theta =
30^{\circ}).
\end{equation}
The elimination of most of the other solutions is based on the Super-K
measurements of the energy spectra during the day and during the
night\cite{fuk11,fuk02}. However, the issue seems not completely settled
yet\cite{str02}.

In summary: Solar neutrinos have yielded strong evidence vor $\nu_e \leftrightarrow
\nu_X \, (X = \mu, \, \tau)$ oscillations. In particular the recent SNO
measurements show explicitly that the solar $\nu$ flux arriving at earth has a
non-$\nu_e$ component. These measurements and their good agreement with the SSM
have solved the long standing solar neutrino problem; they are evidence, in
addition to the results from atmospheric neutrinos, for neutrinos having mass.

\section{Possible neutrino mass schemes}
With two independent $\delta m^2$ values, namely (\ref{e10}) $\delta m^2_{\rm
atm} \approx 3.2 \cdot 10^{-3}$ eV$^2$ and (\ref{e21}) $\delta m^2_{\rm sol}
\approx 5.0 \cdot 10^{-5}$ eV$^2$ one needs three neutrino mass eigenstates
$\nu_{\rm i} = \nu_1, \nu_2, \nu_3$ with masses $m_1, m_2, m_3$ obeying the
relation $\delta m^2_{21} + \delta m^2_{32} + \delta m^2_{13} = 0$ where 
$\delta m^2_{\rm ij} =m^2_{\rm i} - m^2_{\rm j}$. The neutrino flavour
eigenstates $\nu_{\alpha} = \nu_e, \nu_{\mu}, \nu_{\tau}$ are then linear
combinations of the $\nu_{\rm i}$ and vice versa, $\nu_{\alpha} = \sum_{\rm
i}{U_{\alpha \rm i} \nu_{\rm i}}$, in analogy to (\ref{e2}).
\begin{figure}
\begin{center}
\epsfig{file=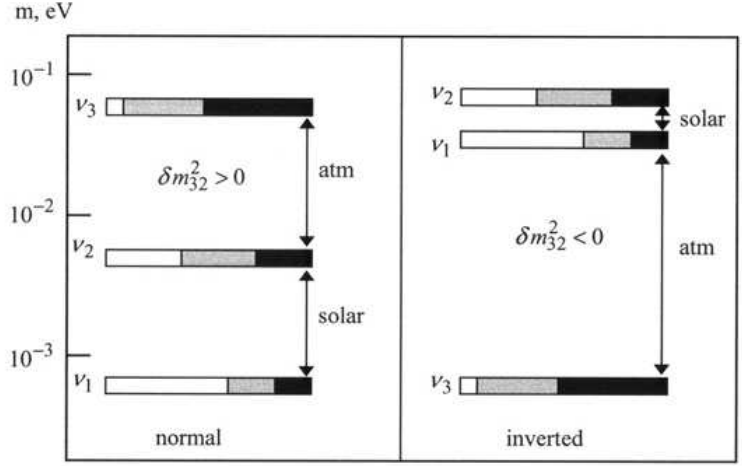,width=10cm}
\end{center}
\caption{Schematic drawing of the normal and inverted hierarchical mass spectrum
of the 3 neutrino mass eigenstates $\nu_{\rm i} \, (i=1, 2, 3)$. The shadings show
the admixtures $\left|U_{\rm ei} \right|^2$ (white), $\left|U_{\mu {\rm i}} 
\right|^2$ (grey) and $\left|U_{\tau {\rm i}} \right|^2$ (black) of the 3
flavour eigenstates $\nu_e, \nu_{\mu}$ and $\nu_{\tau}$, respectively. Adapted
from Ref. \protect\cite{smi01}.}
\label{fig8}
\end{figure}

The absolute neutrino mass scale is still unknown, since a direct measurement
of a neutrino mass has not yet been accomplished. Several possible mass schemes
have been proposed in the literature. The two main categories are:
\begin{list}{--}{\setlength{\leftmargin}{3mm}}
\item
A hierarchical mass spectrum, e.g. $m_1 \ll m_2 \ll m_3$. In this case the
hierarchy may be normal or inverted, as shown in Fig.~\ref{fig8}. If e.g. for the normal
hierarchy one assumes $m_1 \approx 0$, then 
\begin{equation}
\label{e22} 
\begin{array}{lclclclc}
 m_2 & \approx & \sqrt{\delta m^2_{21}} & \approx & \sqrt{\delta m^2_{\rm sol}}
\approx \sqrt{5 \cdot 10^{-5}} \,\, {\rm eV} & \approx & 7 \cdot 10^{-3} \,\,
{\rm eV},\\ [1ex] 
m_3 & \approx & \sqrt{\delta m^2_{31}} & \approx &
\sqrt{\delta m^2_{\rm atm}} \approx \sqrt{3.2 \cdot 10^{-3}} \,\, {\rm eV} &
\approx & 6 \cdot 10^{-2} \,\, {\rm eV}. 
\end{array} 
\end{equation}

\item 
A democratic (nearly degenerate) mass spectrum with $m_1 \approx m_2
\approx m_3 \gg \sqrt{\delta m^2}$. In this case almost any $m_{\nu}$ value
below $\sim 3$ eV (upper limit of $m(\nu_e)$, eq. (\ref{e1})) is possible. In
particular, with $m_{\nu} \sim \mathcal{O}$ (1 eV) neutrinos could contribute noticeably
to the dark matter in the universe. \end{list}

\section*{Acknowledgements} I am grateful to the organizers of the 10.
International Workshop on Multiparticle Production and in particular to Nikos Antoniou, for a very fruitful
and enjoyable meeting with interesting talks and lively discussions on an
island that is famous for its outstanding history and culture as well as for
its beautiful nature. I also would like to thank Mrs. Sybille Rodriguez for her
typing the text and preparing and arranging the figures.


\end{document}